\documentclass[12pt]{article}
\usepackage{graphicx}

\begin{document}

\title{\textbf{Probing high-density behavior of symmetry energy from pion emission in heavy-ion collisions}}

\author{Zhao-Qing Feng$^{a}$\footnote{Corresponding author. Tel. +86 931 4969215. \newline \emph{E-mail address:} fengzhq@impcas.ac.cn},
Gen-Ming Jin$^{b}$}
\date{}
\maketitle

\begin{center}
$^{a}${\small \emph{Institute of Modern Physics, Chinese Academy of
Sciences, Lanzhou 730000, People's Republic of China}}
\end{center}

\textbf{Abstract}
\par
Within the framework of the improved isospin dependent quantum
molecular dynamics (ImIQMD) model, the emission of pion in heavy-ion
collisions in the region 1 A GeV as a probe of nuclear symmetry
energy at supra-saturation densities is investigated systematically,
in which the pion is considered to be mainly produced by the decay
of resonances $\triangle$(1232) and N*(1440). The total pion
multiplicities and the $\pi^{-}/\pi^{+}$ yields are calculated for
selected Skyrme parameters SkP, SLy6, Ska and SIII, and also for the
cases of different stiffness of symmetry energy with the parameter
SLy6. Preliminary results compared with the measured data by the
FOPI collaboration favor a hard symmetry energy of the potential
term proportional to $(\rho/\rho_{0})^{\gamma_{s}}$ with
$\gamma_{s}=2$.

\bigskip

Heavy-ion collisions induced by radioactive beam at intermediate
energies play a significant role to extract the information of
nuclear equation of state (EoS) of isospin asymmetric nuclear matter
under extreme conditions. Besides nucleonic observables such as
rapidity distribution and flow of free nucleons and light clusters
(such as deuteron, triton and alpha etc.), also mesons emitted from
the reaction zone can be probes of the hot and dense nuclear matter.
The energy per nucleon in the isospin asymmetric nuclear matter is
usually expressed as
$E(\rho,\delta)=E(\rho,\delta=0)+E_{\textrm{sym}}(\rho)\delta^{2}+\textsc{O}(\delta^{2})$
in terms of baryon density $\rho=\rho_{n}+\rho_{p}$, relative
neutron excess $\delta=(\rho_{n}-\rho_{p})/(\rho_{n}+\rho_{p})$,
energy per nucleon in a symmetric nuclear matter $E(\rho,\delta=0)$
and bulk nuclear symmetry energy
$E_{\textrm{sym}}=\frac{1}{2}\frac{\partial^{2}E(\rho,\delta)}{\partial
\delta^{2}}\mid_{\delta=0}$. In general, two different forms have
been predicted by some microscopical or phenomenological many-body
approaches. One is the symmetry energy increases monotonically with
density, and the other is the symmetry energy increases initially up
to a supra-saturation density and then decreases at higher
densities. Based on recent analysis of experimental data associated
with transport models, a symmetry energy of the form
$E_{\textrm{sym}}(\rho)\approx 31.6(\rho/\rho_{0})^{\gamma}$ MeV
with $\gamma=0.69-1.05$ was extracted for densities between
0.1$\rho_{0}$ and 1.2$\rho_{0}$ $\cite{Ch05,Li08}$. The symmetry
energy at supra-saturation densities can be investigated by
analyzing isospin sensitive observables in theoretically, such as
the neutron/proton ratio of emitted nucleons, $\pi^{-}/\pi^{+}$,
$\Sigma^{-}/\Sigma^{+}$ and $K^{0}/K^{+}$ $\cite{Li08}$. Recently, a
very soft symmetry energy at supra-saturation densities was pointed
out by fitting the FOPI data $\cite{Re07}$ using IBUU04 model
$\cite{Xi09}$. With the establishment of high-energy radioactive
beam facilities in the world, such as the CSR (IMP in Lanzhou,
China), FAIR (GSI in Darmstadt, Germany), RIKEN (Japan), SPIRAL2
(GANIL in Caen, France) and FRIB (MSU, USA) $\cite{Li08}$, the
high-density behavior of the symmetry energy can be studied more
detail experimentally in the near future. The emission of pion in
heavy-ion collisions in the region 1 A GeV is especially sensitive
as a probe of symmetry energy at supra-saturation densities. Further
investigations of the pion emissions in the 1 A GeV region are still
necessary by improving transport models or developing some new
approaches. The ImIQMD model has been successfully applied to treat
heavy-ion fusion reactions near Coulomb barrier
$\cite{Fe05,Fe08,Wa04}$. Recently, Zhang \emph{et al} analyzed the
neutron-proton spectral double ratios to extract the symmetry energy
per nucleon at sub-saturation density with a similar model
$\cite{Zh08}$. To investigate the pion emission, we further include
the inelastic channels in nucleon-nucleon collisions.

In the ImIQMD model, the time evolutions of the baryons and pions in
the system under the self-consistently generated mean-field are
governed by Hamilton's equations of motion, which read as
\begin{eqnarray}
\dot{\mathbf{p}}_{i}=-\frac{\partial H}{\partial\mathbf{r}_{i}},
\quad \dot{\mathbf{r}}_{i}=\frac{\partial
H}{\partial\mathbf{p}_{i}}.
\end{eqnarray}
Here we omit the shell correction part in the Hamiltonian $H$ as
described in Ref. $\cite{Fe08}$. The Hamiltonian of baryons consists
of the relativistic energy, the effective interaction potential and
the momentum dependent part as follows:
\begin{equation}
H_{B}=\sum_{i}\sqrt{\textbf{p}_{i}^{2}+m_{i}^{2}}+U_{int}+U_{mom}.
\end{equation}
Here the $\textbf{p}_{i}$ and $m_{i}$ represent the momentum and the
mass of the baryons.

The effective interaction potential is composed of the Coulomb
interaction and the local interaction
\begin{equation}
U_{int}=U_{Coul}+U_{loc}.
\end{equation}
The Coulomb interaction potential is written as
\begin{equation}
U_{Coul}=\frac{1}{2}\sum_{i,j,j\neq
i}\frac{e_{i}e_{j}}{r_{ij}}erf(r_{ij}/\sqrt{4L})
\end{equation}
where the $e_{j}$ is the charged number including protons and
charged resonances. The $r_{ij}=|\mathbf{r}_{i}-\mathbf{r}_{j}|$ is
the relative distance of two charged particles.

The local interaction potential is derived directly from the Skyrme
energy-density functional and expressed as
\begin{equation}
U_{loc}=\int V_{loc}(\rho(\mathbf{r}))d\mathbf{r}.
\end{equation}
The local potential energy-density functional reads
\begin{eqnarray}
V_{loc}(\rho)=\frac{\alpha}{2}\frac{\rho^{2}}{\rho_{0}}+
\frac{\beta}{1+\gamma}\frac{\rho^{1+\gamma}}{\rho_{0}^{\gamma}}+
\frac{g_{sur}}{2\rho_{0}}(\nabla\rho)^{2}+\frac{g_{sur}^{iso}}{2\rho_{0}}
[\nabla(\rho_{n}-\rho_{p})]^{2}+     \nonumber \\
\left(a_{sym}\frac{\rho^{2}}{\rho_{0}}+b_{sym}\frac{\rho^{1+\gamma}}{\rho_{0}^{\gamma}}+
c_{sym}\frac{\rho^{8/3}}{\rho_{0}^{5/3}}\right)\delta^{2}+
g_{\tau}\rho^{8/3}/\rho_{0}^{5/3},
\end{eqnarray}
where the $\rho_{n}$, $\rho_{p}$ and $\rho=\rho_{n}+\rho_{p}$ are
the neutron, proton and total densities, respectively, and the
$\delta=(\rho_{n}-\rho_{p})/(\rho_{n}+\rho_{p})$ is the isospin
asymmetry. The coefficients $\alpha$, $\beta$, $\gamma$, $g_{sur}$,
$g_{sur}^{iso}$, $g_{\tau}$ are related to the Skyrme parameters
$t_{0}, t_{1}, t_{2}, t_{3}$ and $x_{0}, x_{1}, x_{2}, x_{3}$
$\cite{Fe08}$. The parameters of the potential part in the symmetry
energy term are also derived directly from Skyrme energy-density
parameters as
\begin{eqnarray}
a_{sym}=-\frac{1}{8}(2x_{0}+1)t_{0}\rho_{0}, \quad b_{sym}=-\frac{1}{48}(2x_{3}+1)t_{3}\rho_{0}^{\gamma}, \nonumber \\
c_{sym}=-\frac{1}{24}\left(\frac{3}{2}\pi^{2}\right)^{2/3}\rho_{0}^{5/3}[3t_{1}x_{1}-t_{2}(5x_{2}+4)].
\end{eqnarray}

The momentum dependent term in the Hamiltonian is the same of the
form in Ref. $\cite{Ai87}$ and expressed as
\begin{equation}
U_{mom}=\frac{\delta}{2}\sum_{i,j,j\neq
i}\frac{\rho_{ij}}{\rho_{0}}[\ln(\epsilon(\textbf{p}_{i}-\textbf{p}_{j})^{2}+1)]^{2},
\end{equation}
with
\begin{equation}
\rho_{ij}=\frac{1}{(4\pi L)^{3/2}}\exp\left[
-\frac{(\textbf{r}_{i}-\textbf{r}_{j})^{2}}{4L}\right],
\end{equation}
which does not distinguish between protons and neutrons. Here the
$L$ denotes the square of the pocket-wave width, which is dependent
on the mass number of the nucleus. The parameters $\delta$ and
$\epsilon$ were determined by fitting the real part of the
proton-nucleus optical potential as a function of incident energy.

In Table 1 we list the ImIQMD parameters related to several typical
Skyrme forces after including the momentum dependent interaction.
The parameters $\alpha$, $\beta$ and $\gamma$ are redetermined in
order to reproduce the binding energy ($E_{B}$=-16 MeV) of symmetric
nuclear matter at saturation density $\rho_{0}$ and to satisfy the
relation $\frac{\partial E/A}{\partial\rho}\mid _{\rho=\rho_{0}}$=0
for a given incompressibility. Combined Eq.(7) with the kinetic
energy part, the symmetry energy per nucleon in the ImIQMD model is
given by
\begin{equation}
E_{sym}(\rho)=\frac{1}{3}\frac{\hbar^{2}}{2m}\left(\frac{3}{2}\pi^{2}\rho\right)^{2/3}+
a_{sym}\frac{\rho}{\rho_{0}}+b_{sym}\left(\frac{\rho}{\rho_{0}}\right)^{\gamma}+
c_{sym}\left(\frac{\rho}{\rho_{0}}\right)^{5/3}.
\end{equation}
More clearly compared with other transport models, the symmetry
energy can be expressed as
\begin{equation}
E_{sym}(\rho)=\frac{1}{3}\frac{\hbar^{2}}{2m}\left(\frac{3}{2}\pi^{2}\rho\right)^{2/3}+
\frac{1}{2}C_{sym}\left(\frac{\rho}{\rho_{0}}\right)^{\gamma_{s}}.
\end{equation}
The value $\gamma_{s}=1$ is used in IQMD model $\cite{Ha98,Ch98}$.
In Fig. 1 we show a comparison of the energy per nucleon in
symmetric nuclear matter with and without the momentum dependent
potentials in the left panel and the nuclear symmetry energy in the
right panel for different cases of Skyrme forces SkP, Sly6, Ska and
SIII from Eq. (10), $\gamma_{s}$=0.5 (soft) and 2 (hard) with
$C_{sym}$=32 MeV in Eq. (11), and also compared with the form
$E_{sym}=31.6(\rho/\rho_{0})^{\mu}$ MeV ($\mu$=0.5 and $\mu$=2)
$\cite{Ch05}$.


Analogously to baryons, the Hamiltonian of pions is represented as
\begin{equation}
H_{\pi}=\sum_{i=1}^{N_{\pi}}\left(\sqrt{\textbf{p}_{i}^{2}+m_{\pi}^{2}}+V_{i}^{Coul}\right),
\end{equation}
where the $\textbf{p}_{i}$ and $m_{\pi}$ represent the momentum and
the mass of the pions. The Coulomb interaction is given by
\begin{equation}
V_{i}^{Coul}=\sum_{j=1}^{N_{B}}\frac{e_{i}e_{j}}{r_{ij}},
\end{equation}
where the $N_{\pi}$ and $N_{B}$ is the total number of pions and
baryons including charged resonances. Thus, the pion propagation in
the whole stage is guided essentially by the Coulomb force. The
in-medium pion potential in the mean field is not considered in the
model. However, the inclusion of the pion optical potential based on
the perturbation expansion of the $\Delta$-hole model gives
negligible influence on the transverse momentum distribution
$\cite{Fu97}$.

The pion is created by the decay of the resonances $\triangle$(1232)
and N*(1440) which are produced in inelastic NN scattering. The
cross section of direct pion production is very small in the
considered energies and not included in the model $\cite{Li01}$. The
reaction channels are given as follows:
\begin{eqnarray}
NN \leftrightarrow N\triangle, & NN \leftrightarrow NN^{\ast}, & NN
\leftrightarrow \triangle\triangle,
\nonumber \\
\Delta \leftrightarrow N\pi, & N^{\ast} \leftrightarrow N\pi.
\end{eqnarray}
The cross sections of each channel to produce resonances are
parameterized by fitting the data calculated with the one-boson
exchange model $\cite{Hu94}$. In the 1 A GeV region, there are
mostly $\Delta$ resonances which disintegrate into a $\pi$ and a
nucleon, however, the $N^{\ast}$ yet gives considerable contribution
to the high energetic pion yield. The energy and momentum dependent
decay width is used in the calculation $\cite{Fe09}$.

Pion meson in heavy-ion collisions is mainly produced at
supra-saturation densities of compressed nuclear matter larger than
the normal density $\rho_{0}$. The production of pions is influenced
by the $\triangle$(1232) and the Fermi motion of baryons in the
vicinity of the threshold energies. The $\pi^{-}$/$\pi^{+}$ ratio is
a sensitive probe to extract the high-density behavior of the
symmetry energy per energy. Shown in Fig. 2 is a comparison of the
measured total pion multiplicity and $\pi^{-}$/$\pi^{+}$ yields by
the FOPI collaboration in central $^{197}$Au+$^{197}$Au collisions
$\cite{Re07}$ and the results calculated by IQMD model $\cite{Ha98}$
as well as by the ImIQMD model for Skyrme parameters SkP, SLy6, Ska
and SIII, which correspond to different modulus of incompressibility
as listed in table 1. The total multiplicity of pion is mainly
determined by the cross sections of the channels $NN \leftrightarrow
N\triangle$. The ImIQMD model with four Skyrme parameters predicts
rather well the total yields at higher incident energies, but
slightly overestimates the values near threshold energies, which may
be influenced by the in-medium cross sections. In this work, we use
the in-vacuum cross sections of nucleon-nucleon elastic and
inelastic collisions. Reasonable consideration of the in-medium
inelastic collisions in producing $\Delta$ and $N^{\ast}$ is still
an open problem in transport models, which have been performed in
Giessen-BUU model $\cite{La01}$. Using the isobar model, one gets
the ratio $\pi^{-}$/$\pi^{+}$=1.95 for pions from the $\Delta$
resonance, and $\pi^{-}$/$\pi^{+}$=1.7 from the $N^{\ast}$ for the
system $^{197}$Au+$^{197}$Au $\cite{St86}$. These relations are
globally valid, i.e. independent of the pion energy. On the other
hand, the statistical model predicts that the $\pi^{-}$/$\pi^{+}$
ratio is sensitive to the difference in the chemical potentials of
neutrons and protons by the relation $\pi^{-}/\pi^{+}\propto \exp
[2(\mu_{n}-\mu_{p})/T]=\exp[8\delta E_{sym}(\rho)/T]$, where the $T$
is nuclear temperature $\cite{Be80}$. The observed energy dependence
of the $\pi^{-}$/$\pi^{+}$ ratio is due to the re-scattering and
absorption process of pions and nucleons in the mean field of the
compressed nuclear matter. We use the free absorption cross sections
in collisions of pions and nucleons by fitting the experimental
data. The branch ratio of the charged $\pi$ and $\pi^{0}$ is
determined by the Clebsch-Gordan coefficients with the decay of the
resonances $\triangle$(1232) and N*(1440). The $\pi^{-}/\pi^{+}$
ratio is sensitive to the stiffness of the symmetry energy at the
lower incident energies. The ImIQMD model can predict the decrease
trend of the $\pi^{-}/\pi^{+}$ ratio with incident energy. While the
ImIQMD model with different Skyrme parameters gives the same
excitation functions of the total pion multiplicity owing to the
same cross sections in the production of pions and resonances for
each case, the $\pi^{-}/\pi^{+}$ yields is different resulting from
the symmetry energy.


The compressed nuclear matter with central density about two times
of the normal density is formed in heavy-ion collisions in the 1 A
GeV region. To extract more information of symmetry energy in
heavy-ion collisions from the pion production, in Fig. 3 we
calculated the time evolution of average central density from low to
high incident energies and the excitation functions of the
$\pi^{-}$/$\pi^{+}$ ratios with the force SLy6, but different
stiffness of the symmetry energy which corresponds to hard
($\gamma_{s}$=2), linear ($\gamma_{s}$=1), soft ($\gamma_{s}$=0.5)
and supersoft (SIII)), and also compared with IQMD results
$\cite{Ha98}$ as well as the FOPI data $\cite{Re07}$. The ImIQMD
model gives larger values of $\pi^{-}$/$\pi^{+}$ than the ones
calculated by IQMD, which mainly results from the cross section of
the channel $N\pi \rightarrow \Delta$ and the larger coefficient
$C_{sym}$. We considered the pion absorption process according to
the Breit-Wigner formula with the cross section given in Ref.
$\cite{Li01}$. Our calculations show that a stiff symmetry energy is
close to experimental data. The results does not support a very soft
symmetry energy at high-density from analyzing the same experimental
data reported in Ref. $\cite{Xi09}$. Situation is different in
IBUU04 model, each nucleon in the evolution is enforced by the
symmetry potential associated with isospin and momentum. Inversely,
a transport model reported in Ref. $\cite{Fe06}$ also predicted the
larger ratios for stiffer symmetry energy from the analysis of the
$\pi^{-}$/$\pi^{+}$ and $K^{0}/K^{+}$ yields. The influence of the
symmetry energy on pion production in heavy-ion collisions is also
studied from the distribution of transverse momentum of the total
charged pions and the ratio $\pi^{-}$/$\pi^{+}$ for the cases of
stiff and soft symmetry energies as shown in Fig. 4. The $\pi^{-}$
mesons are mostly produced from neutron-neutron collisions, and for
a stiff symmetry energy, a wider high-density zone is formed in the
calculation of the ImIQMD model. The larger $\pi^{-}$/$\pi^{+}$
ratio is also clear in the momentum distribution and the larger
errors at the higher transverse momentum are resulted from the
limited simulation events.


The final $\pi^{-}$/$\pi^{+}$ ratio with different stiffness of the
symmetry energy is shown in Fig. 5 as a function of N/Z of the
systems in the reactions $^{40}$Ca+$^{40}$Ca, $^{96}$Ru+$^{96}$Ru,
$^{96}$Zr+$^{96}$Zr and $^{197}$Au+$^{197}$Au, and also plotted the
ratios of N/Z and (N/Z)$^{2}$ as a function of N/Z at incident
energy 0.4A GeV and 1.5A GeV, respectively. The FOPI data
$\cite{Re07}$ and the results calculated by IQMD model $\cite{Ha98}$
are also given for a comparison. Experimental data and calculations
show that an increase trend of the $\pi^{-}$/$\pi^{+}$ ratio in
realistic heavy ion collisions than that predicted by the isobar
model is found at near threshold energy 0.4A GeV, especially for the
larger N/Z systems. The ratio decreases with the incident energy and
the value is located between the lines of (N/Z)$^{2}$ and N/Z at
incident energy 1.5A GeV. The phenomena can be explained from the
fact that the symmetry energy enhances the N/Z ratio in the
high-density region at lower incident energy. The decrease of the
$\pi^{-}$/$\pi^{+}$ ratio with the incident energy is mainly owing
to the production of pions from secondary nucleon-nucleon
collisions, such as a neutron converts a proton by producing
$\pi^{-}$. Subsequent collisions of the energetic proton can convert
again to neutron by producing $\pi^{+}$. One can see that the stiff
symmetry energy is also close to the experimental data. Recently, a
moderately soft symmetry energy with $\gamma_{s}\simeq 0.9\pm0.3$
was extracted from the analysis of neutron-proton elliptic flow of
the FOPI/LAND data for the reaction $^{197}$Au+$^{197}$Au using the
UrQMD model $\cite{Tr09}$. Further experimental works associated
transport models should be performed in more details to get reliable
information of the high-density trend of the symmetry energy in
heavy-ion collisions.


In summary, the pion production in heavy-ion collisions in the
region 1 A GeV is investigated systematically by using the ImIQMD
model. The total multiplicity of produced pion and the
$\pi^{-}/\pi^{+}$ ratio in central collisions are calculated for the
selected Skyrme parameters SkP, SLy6, Ska, SIII which correspond to
different modulus of incompressibility of symmetric nuclear matter
and different cases of the stiffness of symmetry energy, and
compared them with the experimental data by the FOPI collaborations
as well as IQMD results. The $\pi^{-}/\pi^{+}$ excitation functions
for the reaction $^{197}$Au+$^{197}$Au and the dependence of the
$\pi^{-}/\pi^{+}$ ratio on N/Z of reaction systems at energy 0.4A
GeV are compared with the force SLy6, but different stiffness of the
symmetry energy. Calculations show that a stiffer symmetry energy of
the potential term with $\gamma_{s}=2$ is close to the experimental
data.

\textbf{Acknowledgements}

This work was supported by the National Natural Science Foundation
of China under Grant Nos. 10805061 and 10775061, the special
foundation of the president fund, the west doctoral project of
Chinese Academy of Sciences, and major state basic research
development program under Grant No. 2007CB815000.

\newpage
\begin{table}
\caption{ImIQMD parameters and properties of symmetric nuclear
matter for Skyrme effective interactions after the inclusion of the
momentum dependent interaction with parameters $\delta$=1.57 MeV and
$\epsilon$=500 c$^{2}$/GeV$^{2}$} \vspace*{-10pt}
\begin{center}
\def\temptablewidth{0.8\textwidth}
{\rule{\temptablewidth}{1pt}}
\begin{tabular*}{\temptablewidth}{@{\extracolsep{\fill}}ccccccccc}
&Parameters                 &SkM*   &Ska    &SIII   &SVI    &SkP    &RATP   &SLy6 \\
\hline
&$\alpha$ (MeV)             &-325.1 &-179.3 &-128.1 &-123.0 &-357.7 &-250.3 &-296.7 \\
&$\beta$  (MeV)             &238.3  &71.9   &42.2   &51.6   &286.3  &149.6  &199.3 \\
&$\gamma$                   &1.14   &1.35   &2.14   &2.14   &1.15   &1.19   &1.14 \\
&$g_{sur}$(MeV fm$^{2}$)    &21.8   &26.5   &18.3   &14.1   &19.5   &25.6   &22.9 \\
&$g_{sur}^{iso}$(MeV fm$^{2}$)&-5.5 &-7.9   &-4.9   &-3.0   &-11.3  &0.0    &-2.7 \\
&$g_{\tau}$ (MeV)           &5.9    &13.9   &6.4    &1.1    &0.0    &11.0   &9.9 \\
&$C_{sym}$ (MeV)            &30.1   &33.0   &28.2   &27.0   &30.9   &29.3   &32.0 \\
&$a_{sym}$ (MeV)            &62.4   &29.8   &38.9   &42.9   &94.0   &79.3   &130.6 \\
&$b_{sym}$ (MeV)            &-38.3  &-5.9   &-18.4  &-22.0  &-63.5  &-58.2  &-123.7 \\
&$c_{sym}$ (MeV)            &-6.4   &-3.0   &-3.8   &-5.5   &-13.0  &-4.1   &12.8 \\
&$\rho_{\infty}$ (fm$^{-3}$)&0.16   &0.155  &0.145  &0.144  &0.162  &0.16   &0.16 \\
&$m_{\infty}^{\ast}/m$      &0.639  &0.51   &0.62   &0.73   &0.77   &0.56   &0.57 \\
&$K_{\infty}$ (MeV)         &215    &262    &353    &366    &200    &239    &230 \\
\end{tabular*}
{\rule{\temptablewidth}{1pt}}
\end{center}
\end{table}

\begin{figure}
\begin{center}
{\includegraphics*[width=0.8\textwidth]{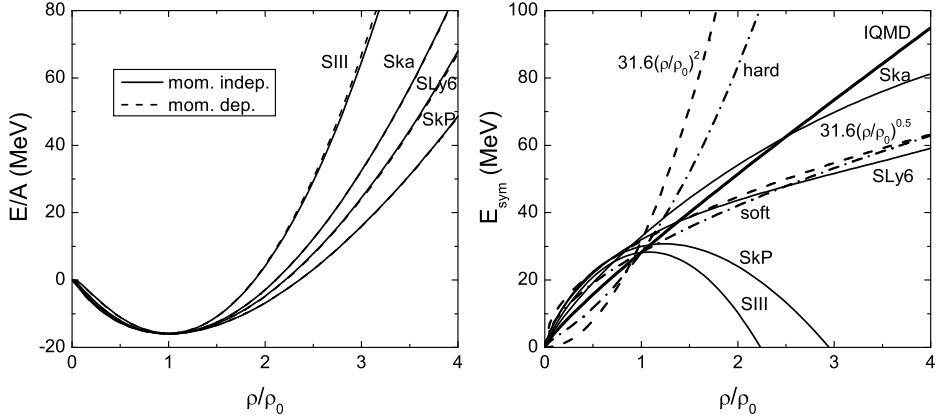}}
\end{center}
\caption{The density dependence of the energy per nucleon in
symmetric nuclear matter at temperature T=0 MeV with and without the
momentum dependent potentials (left panel) and comparison of the
density dependence of the nuclear symmetry energy for different
Skyrme forces SkP, Sly6, Ska and SIII, and the symmetry energy
$E_{sym}=31.6(\rho/\rho_{0})^{\gamma}$ MeV (the two cases
$\gamma$=0.5 and $\gamma$=2) taken in Refs. $\cite{Ch05,Li08}$
(right panel).}
\end{figure}

\begin{figure}
\begin{center}
{\includegraphics*[width=0.8\textwidth]{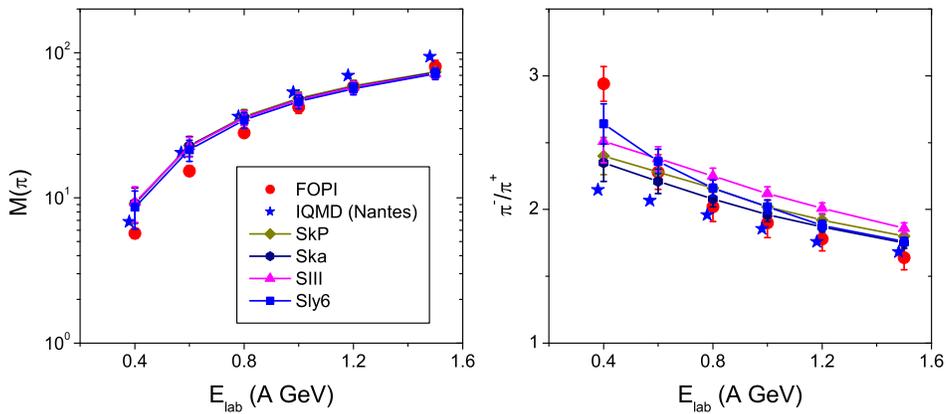}}
\end{center}
\caption{Comparison of calculated pion multiplicity and
$\pi^{-}$/$\pi^{+}$ ratios in central $^{197}$Au+$^{197}$Au
collisions with different Skyrme parameters, and compared with IQMD
results $\cite{Ha98}$ as well as the FOPI data $\cite{Re07}$.}
\end{figure}

\begin{figure}
\begin{center}
{\includegraphics*[width=0.8\textwidth]{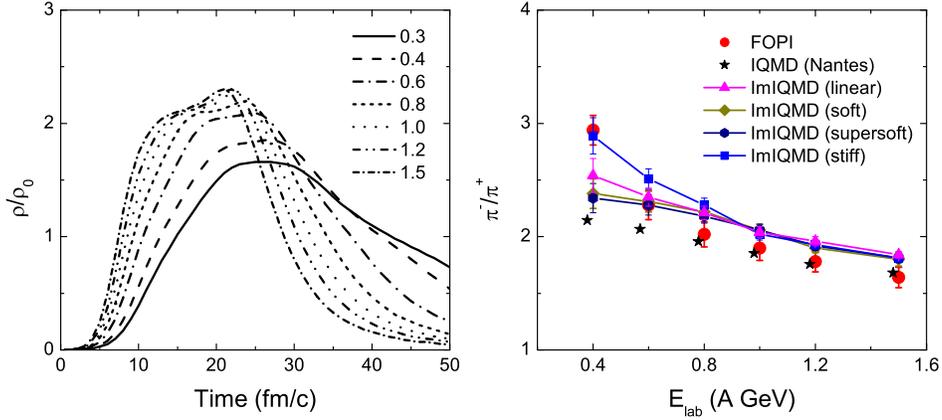}}
\end{center}
\caption{Evolution of average central density at different incident
energies (left panel) and the excitation functions of the
$\pi^{-}$/$\pi^{+}$ ratios at different stiffness of the symmetry
energy (hard, linear, soft and supersoft), and compared with IQMD
results $\cite{Ha98}$ as well as the FOPI data $\cite{Re07}$ (right
panel).}
\end{figure}

\begin{figure}
\begin{center}
{\includegraphics*[width=0.8\textwidth]{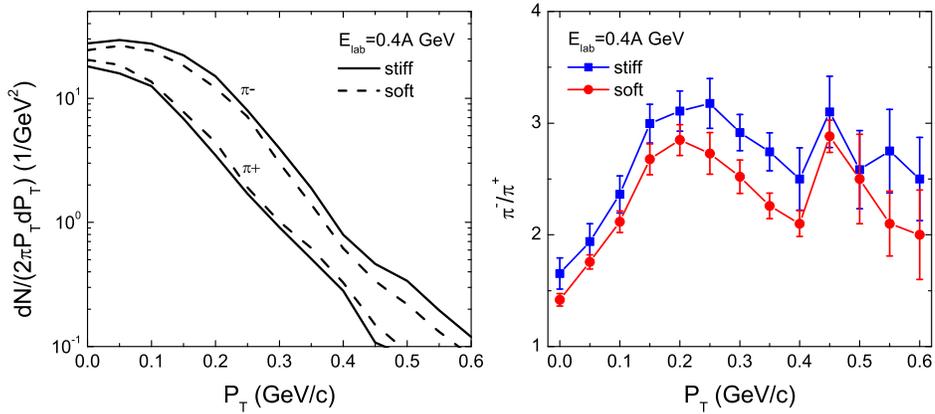}}
\end{center}
\caption{Distributions of transverse momentum of final $\pi^{-}$ and
$\pi^{+}$ and the ratio $\pi^{-}$/$\pi^{+}$ for the cases of stiff
and soft symmetry energies in the reaction $^{197}$Au+$^{197}$Au at
incident energy $E_{lab}=$ 0.4A GeV.}
\end{figure}

\begin{figure}
\begin{center}
{\includegraphics*[width=0.8\textwidth]{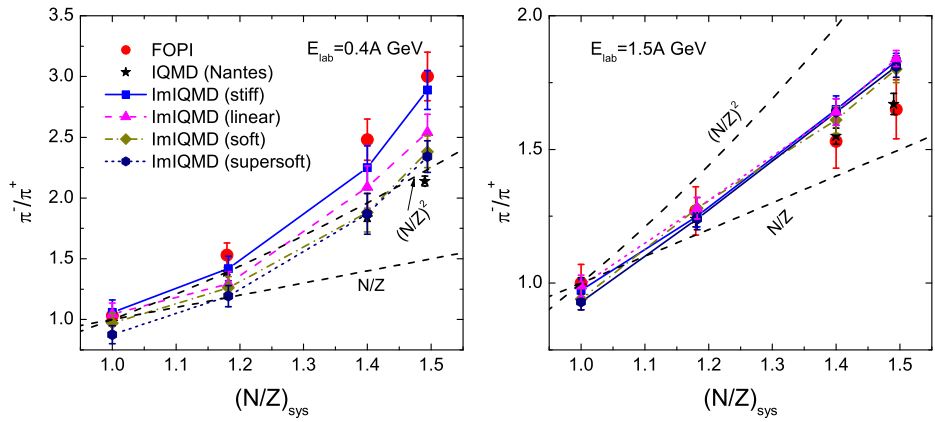}}
\end{center}
\caption{The $\pi^{-}$/$\pi^{+}$ yields as a function of the neutron
over proton N/Z of reaction systems for head on collisions at
incident energy $E_{lab}=$ 0.4A GeV and 1.5A GeV, respectively.}
\end{figure}

\end{document}